\documentclass[letter]{jpsj3}

\title{Superparamagnetism induced by polar nanoregions in relaxor ferroelectric (1$-$$x$)BiFeO$_{3}$-$x$BaTiO$_{3}$}

\author{Minoru Soda$^{1,2}$, Masato Matsuura$^{3}$, Yusuke Wakabayashi$^{2}$, and Kazuma Hirota$^{4}$ %\\
}
\inst{$^{1}$Neutron Science Laboratory, Institute for Solid State Physics, University of Tokyo, Kashiwa, Chiba 277-8581, Japan\\
$^{2}$Division of Materials Physics, Graduate School of Engineering Science, Osaka University, Toyonaka, Osaka 560-8531, Japan\\
$^{3}$Institute for Materials Research, Tohoku University, Katahira, Aoba-ku, Sendai 980-8577, Japan\\
$^{4}$Department of Earth and Space Science, Graduate School of Science, Osaka University, Toyonaka, Osaka 560-0043, Japan %\\
}
\abst{A new class of superparamagnetism was found in relaxor ferroelectric 2/3BiFeO$_{3}$-1/3BaTiO$_{3}$. The size of the magnetic particle, estimated from the superparamagnetic magnetization curve, coincides with the size of the polar nanoregion (PNR), which governs the relaxor ferroelectric property. This suggests that the magnetic domain is identical to the PNR. The temperature variations in the sizes of the magnetic domains and PNRs estimated by our neutron diffraction measurements support this picture. Since the same domain provides both electric and magnetic properties, strong coupling between the two properties through the domain size is expected.}

\kword{superparamagnetism, relaxor, polar nanoregion, (1$-$$x$)BiFeO$_{3}$-$x$BaTiO$_{3}$}

\begin{document}
\maketitle

Recently, multiferroic systems, in which magnetic and ferroelectric orders coexist, have been extensively investigated both experimentally and theoretically.\cite{Fiebig1,Eerenstein} In these multiferroic systems, the application of a magnetic field $H$ causes a change in electric polarization $P$, namely, magnetoelectric (ME) effects.\cite{Kimura} Although multiferroic systems have already been classified into several groups in terms of the origins of their ME effects,\cite{Cheong} it is also important to explore other origins of ME effects or the correlation between the dielectric property and magnetism. In some cases, the electric and magnetic domain structures interact with each other.\cite{Fiebig,Kishimoto} Inversely, macroscopic ME coupling can be modulated by their domain structures. Here, we focused on a relaxor ferroelectric system, which involves ferroelectric nano-domains,\cite{Bokov,Burns} (1$-$$x$)BiFeO$_{3}$-$x$BaTiO$_{3}$(BFO-$x$BTO); this system is a rare example of relaxor ferroelectrics having magnetic ions.

In relaxor systems, the temperature dependence of the dielectric permittivity ($\epsilon$) shows a broad maximum and a large frequency dependence.\cite{Bokov,Cross} To explain the physical behaviors of relaxors, the existence of randomly oriented very local polar regions, or polar nanoregions (PNRs), that start to appear from a much higher temperature ($T_{\mathrm{d}}$) than the temperature showing maximum $\epsilon$, was proposed several decades ago.\cite{Burns} Such nanoregions have been observed in various compounds as butterfly-shaped diffuse scatterings.\cite{Matsuura,Xu} The dielectric permittivity of BFO-$x$BTO around $x$$\sim$1/3 also shows a relaxor-like behavior,\cite{Kumar} although both rhombohedral BiFeO$_{3}$ ($x$=0)\cite{Sosnowska} and tetragonal BaTiO$_{3}$ ($x$=1)\cite{Merz} exhibit ordinary ferroelectricity. The concentration $x$=1/3 is close to the morphotropic phase boundary (MPB),\cite{Kumar,Yoneda,Chandarak} at which the dielectric constant is highest for typical relaxors\cite{Ohwa}. Recent studies of the BFO-$x$BTO system, microscopic \cite{Yoneda} and macroscopic \cite{Chandarak} studies around the MPB and the butterfly-shaped diffuse scatterings \cite{Ozaki}, make the relaxor properties of BFO-$x$BTO clear. In the present study, the interaction between the PNR and magnetism was studied through both macroscopic property measurements and microscopic neutron-scattering measurements on single crystalline samples of BFO-$x$BTO. Our results with BFO-$x$BTO indicate that BFO-1/3BTO has PNR-sized nano magnetic domains, and their moments cause superparamagnetic behavior.
\begin{figure}[t]
\begin{center}\leavevmode
\includegraphics[width=8 cm]{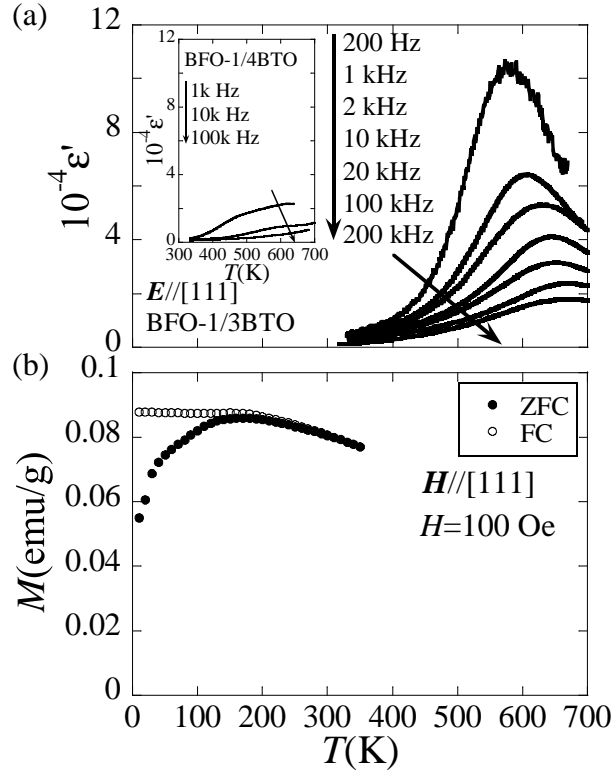}
\caption{
(a) $T$-dependence of $\epsilon'$ taken at several frequencies in BFO-1/3BTO. The inset shows the $T$-dependence of $\epsilon'$ in BFO-1/4BTO. (b) $T$-dependence of the magnetization of BFO-1/3BTO taken at $H$=100 Oe with conditions of zero-field cooling (ZFC) and field cooling (FC).
}
\label{Fig. 1}
\end{center}
\end{figure}

Single crystals of BFO-1/3BTO, which is at the MPB, and of BFO-1/4BTO, which is separated from the MPB, were grown by the floating zone method. The crystals were checked by powder X-ray diffraction to ensure they did not have significant impurity phases. Since no report has been made for single crystals of BFO-$x$BTO, we measured the dielectric permittivity and the magnetization $M$ using an $LCR$ meter and a Quantum Design SQUID magnetometer, respectively, to clarify the macroscopic properties. The electric properties of BFO-$x$BTO polycrystalline samples were reported as relaxor ferroelectrics only in the vicinity of the MPB.\cite{Kumar} The temperature ($T$) dependences of the real part of the dielectric constant $\epsilon'$ for the single crystals of $x$=1/3 and 1/4 taken at various frequencies with electric field {\itshape\bfseries{E}}//[111] are shown in Fig. 1(a); throughout this paper, we use the pseudo-cubic unit cell with the lattice parameter 4.0 \AA. As expected, the results for $x$=1/3 exhibit a relaxor characteristic, while BFO-1/4BTO does not show such properties. The dielectric property of the single crystals is similar to the properties of the polycrystalline sample.\cite{Kumar} In BFO-1/3BTO, the activation energy in a Vogel-Fulcher law estimated from the dielectric-peak temperature and the frequency was 0.12$\pm$0.05 eV, which agrees with the typical relaxor\cite{Zhao}.

As for the magnetization of the polycrystalline samples, weak ferromagnetism was reported for a wide $x$-range.\cite{Kumar3} The magnetic field ($H$)-dependences of $M$ with {\itshape\bfseries{H}}//[111] for the single crystal of BFO-1/3BTO are shown in Fig. 2. The magnetization saturates with small external $H$, which agrees qualitatively with the polycrystalline property.
\begin{figure}[t]
\begin{center}\leavevmode
\includegraphics[width=8 cm]{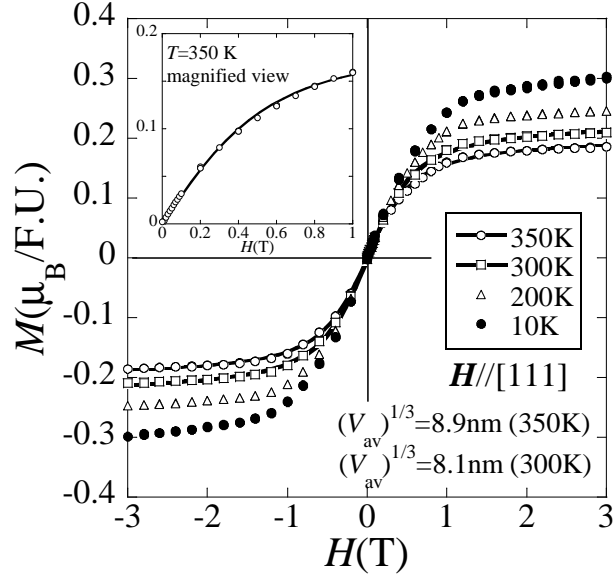}
\caption{
$H$-dependence of $M$ of the single crystal of BFO-1/3BTO measured at several temperatures. Above 300 K, the curves obtained by fitting with the Langevin function are shown by the solid lines. The inset shows the magnified view at 350 K in the low-$H$ region.
}
\label{Fig. 2}
\end{center}
\end{figure}

For the samples cut from $x$=1/3 and 1/4 single crystals used in the measurements of the dielectric constant and the magnetism, we carried out detailed neutron studies. Neutron measurements were carried out using the ISSP triple axis spectrometer PONTA and HQR installed at JRR-3 of JAEA, Japan. The crystal was oriented with the [110] and [001] axes in the scattering plane.

For both $x$=1/3 and 1/4, magnetic reflections at the {\itshape\bfseries{Q}}-points ($\frac{h'}{2}$,$\frac{h'}{2}$,$\frac{l'}{2}$) ($h'$ and $l'$ are odd) were observed at low temperatures. We started by determining the magnetic structure of BFO-1/3BTO. Assuming the magnitudes of all Fe-moments are equal, we found that the magnetic structure of BFO-1/3BTO is the so-called G-type\cite{Wollan} antiferromagnetic one. The inset of Fig. 3(a) shows the integrated intensities of the magnetic reflections collected at 150 K ($I_{\mathrm{obs}}$) against those of the G-type anitiferromagnetic structure ($I_{\mathrm{cal}}$) calculated for five superlattice points. The excellent agreement between the experimental results and the calculated intensity confirms that our analysis is correct. The magnitude of the G-type antiferromagnetic Fe-moments is 4.5$\pm$0.2 $\mu_{\mathrm{B}}$. This means that the Fe$^{3+}$ ions have nearly full magnetic moments at low temperatures, and the observed magnetization shown in Fig. 2 is caused by a cant in the magnetic moments by 5.5$^\circ$$\pm$1$^\circ$; this amount of ferromagnetic moment is too small for neutron scattering to be observed. The cant angle is nearly constant below 350 K. The magnetic structure of BFO-1/4BTO was found to be the same as that of BFO-1/3BTO from similar measurements.
\begin{figure}[t]
\begin{center}\leavevmode
\includegraphics[width=8 cm]{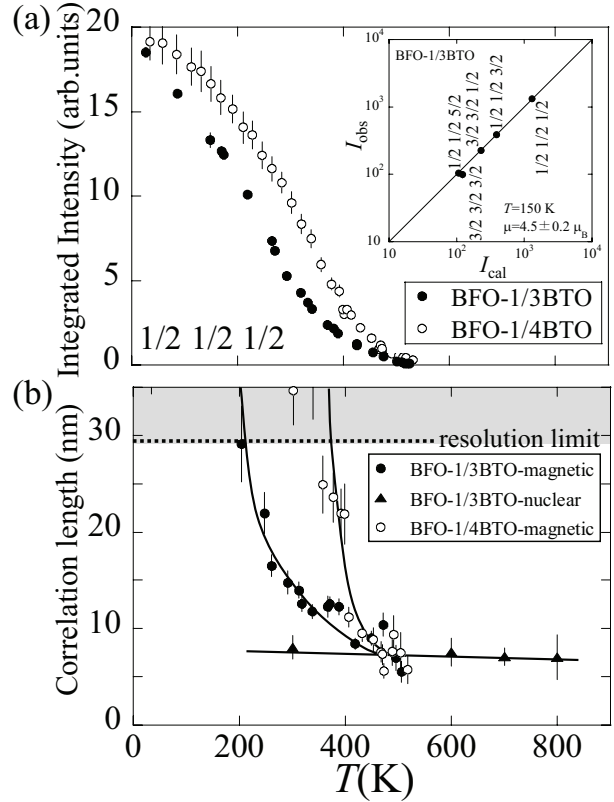}
\caption{
(a) The $T$-dependences of the integrated intensities estimated from 1/2 1/2 1/2 magnetic reflections of BFO-1/3BTO (closed circles) and BFO-1/4BTO (open circles). The inset shows the integrated intensities of the magnetic reflections collected at 150 K in BFO-1/3BTO plotted against those calculated using the G-type magnetic structure. (b) Temperature dependence of the correlation lengths. Closed and open circles indicate the magnetic correlation lengths of BFO-1/3BTO and BFO-1/4BTO, respectively. Closed triangles indicate the nuclear correlation length estimated by nuclear diffuse scattering.
}
\label{Fig. 3}
\end{center}
\end{figure}

Although the system has the simple G-type magnetic structure, the $T$-dependence of the magnetic reflection is not simple. The $T$-dependences of the integrated intensity and the magnetic correlation length estimated from 1/2 1/2 1/2 magnetic peak profiles are shown in Figs. 3(a) and 3(b), respectively. In BFO-1/4BTO, which is separated from the MPB, an antiferromagnetic order shows the long-range one in a wide $T$-region. As $T$ increases, the long-ranged antiferromagnetic order disappears straightforwardly with a rather narrow temperature region of a short-ranged-ordered state. On the other hand, the $T$-dependence of the magnetic reflection in BFO-1/3BTO, which is at the MPB, indicates that the magnetic correlation is not long-ranged in a wide temperature region, 200 K to 500 K. Around 500 K, the magnetic correlation length approaches 8 nm as $T$ increases. This behavior is unusual because many short-range orderings are observed within a narrow temperature region, ($T-T_{\mathrm{c}}$)/$T_{\mathrm{c}}$$<$0.2\cite{Hirota, Zimmermann}, where $T_{\mathrm{c}}$ is the transition temperature.

To examine the dielectric properties, nuclear diffuse scattering was measured. The PNR, which is related to the relaxor property, induces nuclear diffuse scattering for typical relaxors.\cite{Matsuura,Xu} Figures 4(a) and 4(b) show the diffuse intensity distribution around the 112 Bragg reflection at 600 K for BFO-1/3BTO and BFO-1/4BTO, respectively. Strong nuclear diffuse scattering was observed only in BFO-1/3BTO, which has a large dielectric constant. The nuclear correlation length, estimated from the profile width of the nuclear diffuse scattering along the direction perpendicular to the scattering vector, is plotted in Fig. 3(b) against $T$ with closed triangles. The nuclear correlation length is about 8 nm in a wide temperature region. This size is very similar to the PNR observed by transmission electron microscopy (TEM) in this system.\cite{Ozaki} Since there is no other known structure having this spatial-length scale in this system, we conclude that the nuclear diffuse scattering is given by the PNR. Figure 4(c) shows the temperature dependence of the intensity at (1$\pm$$\delta$,1$\pm$$\delta$,2$\mp$$\delta$) with $\delta$=0.03$\sim$0.06 in BFO-1/3BTO. Since the intensity of the nuclear diffuse scattering decreases as $T$ increases, it does not originate from the lattice vibration. The nuclear diffuse scattering quickly decreases above 600 K, where the value of $\epsilon'$ peaks, and vanishes at 800 K as $T$ increases, and thus $T_{\mathrm{d}}$ is found to be 800 K. The little temperature variation of the nuclear diffuse scattering profile below $T_{\mathrm{d}}$ appears to be different from ordinary relaxor systems\cite{Xu2}; we need a systematic measurement involving different scattering plane to clarify this difference, which is out of scope of this study. In the rest of this paper, we discuss the relation between the magnetism and the nano structure observed in the nuclear diffuse scattering measured in the ($HHL$)-scattering plane.

\begin{figure}[t]
\begin{center}\leavevmode
\includegraphics[width=15 cm]{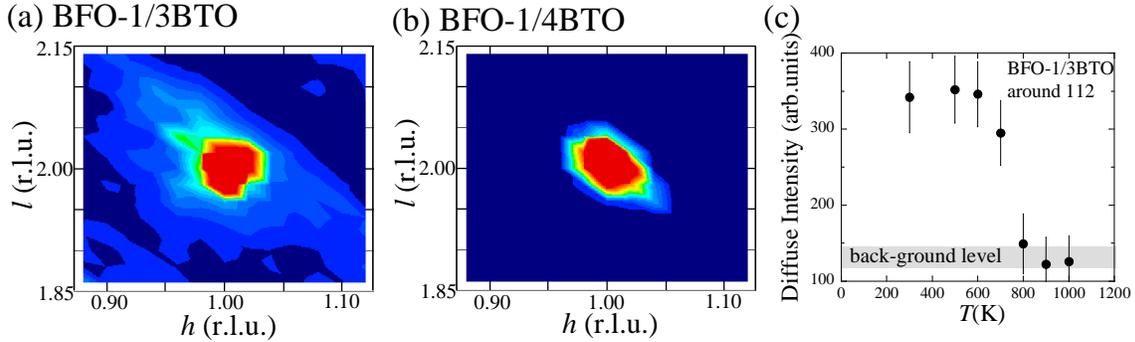}
\caption{
(Color online) Contour plots of the diffuse intensities at 600 K around the 112 reflection for the scattering plane ($h$$h$$l$) in (a) BFO-1/3BTO and (b) BFO-1/4BTO. (c) Temperature dependence of the intensity of the diffuse scattering around the 112 reflection in BFO-1/3BTO.
}
\label{Fig. 4}
\end{center}
\end{figure}
From these neutron results, we propose that the G-type antiferromagnetic order is suppressed by PNRs and their domain walls, which are associated with the randomness of the crystal structure. The PNR size estimated by the nuclear diffuse scattering is 8 nm, and the magnetic correlation length also approaches 8 nm at around 500 K as $T$ increases. These results mean that the magnetic correlation is limited within a PNR above 200 K. As $T$ decreases, the magnetic correlation between neighboring PNRs begins to grow gradually. Since PNRs and their domain walls are the origins of the short-range order, significant suppression of the antiferromagnetic order was observed in BFO-1/3BTO, which has PNR.

According to this interpretation, the $x$=1/3 system contains a vast number of nano magnetic particles. This structure should induce superparamagnetism. We reexamined the magnetization curve in order to confirm our microscopic interpretation. The $H$-dependence of $M$ at 350 K, whose residual magnetization was less than that of our experimental sensitivity, was fitted perfectly by the Langevin function with 5.3$\times$10$^{19}$ moments/mol of 2.3$\times$10$^{3}$ $\mu_{\mathrm{B}}$, as shown in Fig. 2. Similarly, the $M$-$H$ curve at 300 K is also reproduced by the Langevin function. These ensure that BFO-1/3BTO has the superparamagnetic property above 300 K. The ferromagnetic moment of one Fe$^{3+}$ and the average volume of the magnetic domain ($V_{\mathrm{av}}$=total volume/domain number) obtained by the fitting are 0.30 $\mu_{\mathrm{B}}$ and (8.9 nm)$^{3}$, respectively, at $T$=350 K. The domain size is the same as the PNR size, showing that the origin of the superparamagnetism is the nano-magnetic domain induced by the PNR.

At a low temperature, the magnetization deviates from the Langevin function. The residual magnetization at 10 K estimated from Fig. 1(b) was 0.003 $\mu_{\mathrm{B}}$/F.U., while no residual magnetization is expected in the Langevin function. This deviation decreases as $T$ increases, and vanishes at 200 K. This characteristic temperature is the blocking temperature ($T_{\mathrm{B}}$) in superparamagnetism.\cite{McHenry,Fonseca} Below $T_{\mathrm{B}}$, the paramagnetic behavior is blocked. In an ordinary superparamagnet having fixed-sized magnetic particles, this blocking is caused by the decrease in the thermal fluctuation. In addition, the growth of the magnetic correlation length with decreasing temperature can contribute to the blocking in this system. The longer correlation length makes the ``magnetic particle'' larger, and the magnetic correlation length increases significantly around 200 K, which coincides with the blocking temperature.

Let us reexamine the results with an $x$=1/4 sample by following the case for $x$=1/3. The residual magnetization was larger than 0.001 $\mu_{\mathrm{B}}$/F.U. at 350 K, the highest temperature our magnetometer can reach. This result means $T_{\mathrm{B}}$$>$350 K. Higher $T_{\mathrm{B}}$ implies a larger ``magnetic particle,'' which may be given by the long magnetic correlation length at this temperature (see Fig. 3(b)) or by the long correlation length of the lattice caused by the lack of PNRs (see Fig. 4(b)). Therefore, the results for $x$=1/4 are qualitatively understood based on the microscopic picture constructed for $x$=1/3.

Finally, the microscopic scenario of the relationship between magnetism and the rhombohedrally distorted PNR is discussed. In the polar region, the inversion symmetry is broken and Dzyaloshinskii-Moriya (DM) interaction works. Here, we take one example of a rhombohedral iron oxide: hematite\cite{Dzyaloshinskii}. In hematite, DM interaction makes a small cant moment perpendicular to the three-fold axis. This relation is a result of symmetry, and the same relation is also expected in a PNR in a BFO-1/3BTO system. The cant moment should be perpendicular to the [111] axis, the polarization direction. The in-plane freedom of the magnetic ordering direction as well as the eight equivalent $\langle$111$\rangle$ directions allow the magnetic flexibility required for superparamagnetism.

In conclusion, both microscopic and macroscopic studies of single crystalline samples of (1$-$$x$)BiFeO$_{3}$-$x$BaTiO$_{3}$ were carried out to examine the relationship between the relaxor-like dielectric property and magnetism. From the neutron results, the PNR was found to affect the magnetic ordering significantly, and the resulting magnetic nano-domains are the new origin of superparamagnetism. Superparamagnetism is very unique because of its inherent connection between dielectric and magnetic properties.

We are grateful to Prof. T. Kimura, Prof. S. Mori, and Mr. T. Ozaki for their fruitful discussions. Work at JRR-3 was supported by ISSP of the University of Tokyo. This work was supported by KAKENHI (19052002 and 20740171) and the Global COE program (G10).

\end{document}